\documentclass{article}

\usepackage{microtype}
\usepackage{graphicx}
\usepackage{booktabs} 

\usepackage{hyperref}

\usepackage[accepted]{icml2021_style/icml2021}

\usepackage{hhline}
\usepackage{appendix}
\usepackage{amsmath}
\usepackage{amsfonts}
\usepackage{amssymb}
\usepackage{amsthm}
\usepackage{array}
\usepackage{multirow}
\usepackage{caption}
\usepackage{subcaption}

\usepackage{enumitem}
\setlist{nosep}

\usepackage{booktabs}
\usepackage{makecell}
\usepackage{tikz}
\usepackage{pdfpages}

\newcommand{\superhard}{\textit{Super~Hard}}
\newcommand{\ultrahard}{\textit{Ultra~Hard}}

\newcommand{\statespace}{\mathbb{S}}
\newcommand{\state}{s}
\newcommand{\timespace}{\mathbb{T}}
\newcommand{\horizon}{\mathrm{T}}
\newcommand{\timestep}{t}
\newcommand{\jointactionspace}{\mathbb{U}}
\newcommand{\jointaction}{\mathbf{u}}
\newcommand{\action}{u}
\newcommand{\statetransitionfunction}{P}
\newcommand{\rewardfunction}{R}
\newcommand{\reward}{r}
\newcommand{\jointobservationspace}{\mathbb{O}}
\newcommand{\jointobservation}{\mathbf{o}}
\newcommand{\observation}{o}
\newcommand{\jointobservationhistoryspace}{\mathbb{H}}
\newcommand{\jointobservationhistory}{\mathbf{h}}
\newcommand{\observationhistory}{h}
\newcommand{\observationfunction}{O}
\newcommand{\agentspace}{\mathbb{K}}
\newcommand{\agentcounter}{k}
\newcommand{\numberofagents}{\mathrm{K}}
\newcommand{\discountfactor}{\gamma}
\newcommand{\policy}{\pi}
\newcommand{\tderror}{\delta}

\newcommand{\huberloss}{\mathcal{L}_\kappa}
\newcommand{\asymmetrichuberloss}{\rho}
\newcommand{\loss}{\mathcal{L}}
\newcommand{\utilityexp}{Q}
\newcommand{\utility}{Z}
\newcommand{\joint}{\mathrm{jt}}
\newcommand{\quantilefunction}{F^{-1}}

\newcommand{\quantile}{\omega}
\newcommand{\pdf}{f}
\newcommand{\cdf}{F}
\newcommand{\pdfmodelparameter}{\alpha}
\newcommand{\modelparameter}{\beta}
\newcommand{\inversecdf}{F^{-1}}
\newcommand{\numberofquantiles}{\mathrm{N}}
\newcommand{\numberofquantilesamples}{\mathrm{N}'}
\newcommand{\numberofquantiletestsamples}{\hat{\mathrm{N}}}
\newcommand{\implicitquantilefunction}{g}
\newcommand{\stateembeddingfunction}{\psi}
\newcommand{\cosineembeddingfunction}{\phi}
\newcommand{\cosinedimension}{\mathrm{n}}
\newcommand{\monotonicfunction}{M}
\newcommand{\meandecompositionfunction}{\Psi}
\newcommand{\shapedecompositionfunction}{\Phi}
\newcommand{\additionalparameterfunction}{\Lambda}
\newcommand{\additivity}{\textit{Additivity}}
\newcommand{\monotonicity}{\textit{Monotonicity}}
\newcommand{\diql}{\textit{DIQL}}
\newcommand{\ddn}{\textit{DDN}}
\newcommand{\dmix}{\textit{DMIX}}

\newcommand{\digm}{\textit{DIGM}}

\definecolor{myblue}{HTML}{0000B5}
\definecolor{crimson}{HTML}{B30000}
\definecolor{indigo}{HTML}{4B0082}
\newcommand{\bb}[1]{\textcolor{myblue}{#1}}
\newcommand{\cc}[1]{\textcolor{crimson}{#1}}
\newcommand{\pp}[1]{\textcolor{indigo}{#1}}

\newcounter{conjecture}
\newcounter{proposition}
\newcounter{lemma}
\newcounter{corollary}
\newcounter{example}
\newcounter{definition}
\newcounter{remark}

\newtheorem{theorem}{Theorem}[section]

\newtheorem{definition}[definition]{Definition}

\icmltitlerunning{DFAC Framework: Factorizing the Value Function via Quantile Mixture for Multi-Agent Distributional Q-Learning}

\begin{document}

\twocolumn[
\icmltitle{DFAC Framework: Factorizing the Value Function via \\
Quantile Mixture for Multi-Agent Distributional Q-Learning}



\icmlsetsymbol{equal}{*}
\begin{icmlauthorlist}
\icmlauthor{Wei-Fang Sun}{nthu,nv,intern}
\icmlauthor{Cheng-Kuang Lee}{nv}
\icmlauthor{Chun-Yi Lee}{nthu}
\end{icmlauthorlist}
\icmlaffiliation{nthu}{Department of Computer Science, National Tsing Hua University, Taiwan}
\icmlaffiliation{nv}{NVIDIA AI Technology Center, NVIDIA Corporation}
\icmlaffiliation{intern}{Wei-Fang Sun contributed to the work during his NVIDIA internship}
\icmlcorrespondingauthor{Chun-Yi Lee}{cylee@cs.nthu.edu.tw}

\icmlkeywords{Reinforcement Learning, Multi-Agent RL, Distributional RL}

\vskip 0.3in
]



\printAffiliationsAndNotice{}  

\begin{abstract}
In fully cooperative multi-agent reinforcement learning (MARL) settings, the environments are highly stochastic due to the partial observability of each agent and the continuously changing policies of the other agents. 
To address the above issues, we integrate distributional RL and value function factorization methods by proposing a \textbf{D}istributional Value Function \textbf{Fac}torization (DFAC) framework to generalize expected value function factorization methods to their DFAC variants. 
DFAC extends the individual utility functions from deterministic variables to random variables, and models the quantile function of the total return as a quantile mixture. 
To validate DFAC, we demonstrate DFAC's ability to factorize a simple two-step matrix game with stochastic rewards and perform experiments on all \superhard{} tasks of StarCraft Multi-Agent Challenge, showing that DFAC is able to outperform expected value function factorization baselines.
\end{abstract}



%


\setlength{\belowdisplayskip}{4pt} \setlength{\belowdisplayshortskip}{4pt}
\setlength{\abovedisplayskip}{4pt} \setlength{\abovedisplayshortskip}{4pt}

\section{Introduction}
\label{sec:introduction}

In multi-agent reinforcement learning (MARL), one of the popular research directions is to enhance the training procedure of fully cooperative and decentralized agents. Examples of such agents include a fleet of unmanned aerial vehicles (UAVs), a group of autonomous cars, etc. This research direction aims to develop a decentralized and cooperative behavior policy for each agent, and is especially difficult for MARL settings without an explicit communication channel. The most straightforward approach is independent Q-learning (IQL)~\cite{Tan1993IQL}, where each agent is trained independently, with their behavior policies aimed to optimize the overall rewards in each episode. Nevertheless, each agent's policy may not converge owing to two main difficulties: (1) non-stationary environments caused by the changing behaviors of the agents, and (2) spurious reward signals originated from the actions of the other agents. The agent’s partial observability of the environment further exacerbates the above issues.
Therefore, in the past few years, a number of MARL researchers turned their attention to centralized training with decentralized execution (CTDE) approaches, with an objective to stabilize the training procedure while maintaining the agents' abilities for decentralized execution~\cite{Oliehoek2016CTDE}. Among these CTDE approaches, value function factorization methods~\cite{Sunehag2018VDN,Rashid2018QMIX,Son2019QTRAN} are especially promising in terms of their superior performances and data efficiency~\cite{Samvelyan2019SMAC}.


Value function factorization methods introduce the assumption of individual-global-max (IGM)~\cite{Son2019QTRAN}, which assumes that each agent's optimal actions result in the optimal joint actions of the entire group.
Based on IGM, the total return of a group of agents can be factorized into separate utility functions~\cite{Guestrin2001Utility} (or simply `\textit{utility}' hereafter) for each agent. The utilities allow the agents to independently derive their own optimal actions during execution, and deliver promising performance in StarCraft Multi-Agent Challenge (SMAC)~\cite{Samvelyan2019SMAC}. Unfortunately, current value function factorization methods only concentrate on estimating the expectations of the utilities, overlooking the additional information contained in the full return distributions. Such information, nevertheless, has been demonstrated beneficial for policy learning in the recent literature~\cite{Lyle2019Comparative}.


In the past few years, distributional RL has been empirically shown to enhance value function estimation in various single-agent RL (SARL) domains~\cite{Bellemare2017C51,Dabney2018QR-DQN,Dabney2018IQN,Rowland2019ER-DQN,Yang2019FQF}. Instead of estimating a single scalar Q-value, it approximates the probability distribution of the return by either a categorical distribution~\cite{Bellemare2017C51} or a quantile function~\cite{Dabney2018QR-DQN,Dabney2018IQN}. Even though the above methods may be beneficial to the MARL domain due to the ability to capture uncertainty, it is inherently incompatible to expected value function factorization methods (e.g., value decomposition network (VDN)~\cite{Sunehag2018VDN} and QMIX~\cite{Rashid2018QMIX}). The incompatibility arises from two aspects: (1) maintaining IGM in a distributional form, and (2) factorizing the probability distribution of the total return into individual utilities. As a result, an effective and efficient approach that is able to solve the incompatibility is crucial and necessary for bridging the gap between value function factorization methods and distributional RL.


In this paper, we propose a \textbf{D}istributional Value Function \textbf{Fac}torization (DFAC) framework, to efficiently integrate value function factorization methods with distributional RL.
DFAC solves the incompatibility by two techniques: (1) Mean-Shape Decomposition and (2) Quantile Mixture. The former allows the generalization of expected value function factorization methods (e.g., VDN and QMIX) to their DFAC variants without violating IGM. The latter allows the total return distribution to be factorized into individual utility distributions in a computationally efficient manner.
To validate the effectiveness of DFAC, we first demonstrate the ability of distribution factorization on a two-step matrix game with stochastic rewards. Then, we perform experiments on all \superhard{} maps in SMAC. The experimental results show that DFAC offers beneficial impacts on the baseline methods in all \superhard{} maps. In summary, the primary contribution is the introduction of DFAC for bridging the gap between distributional RL and value function factorization methods efficiently by mean-shape decomposition and quantile mixture.

\section{Background and Related Works}
\label{sec:background}

In this section, we introduce the essential background material for understanding the contents of this paper. We first define the problem formulation of cooperative MARL and CTDE. Next, we describe the conventional formulation of IGM and the value function factorization methods. Then, we walk through the concepts of distributional RL, quantile function, as well as quantile regression, which are the fundamental concepts frequently mentioned in this paper. After that, we explain the implicit quantile network, a key approach adopted in this paper for approximating quantiles.  Finally, we bring out the concept of quantile mixture, which is leveraged by DFAC for factorizing the return distribution.


\subsection{Cooperative MARL and CTDE}
\label{subsec:background_cooperative_marl_and_ctde}

In this work, we consider a fully cooperative MARL environment modeled as a decentralized and partially observable Markov Decision Process (Dec-POMDP)~\cite{Oliehoek2016DECPOMDP} with stocastic rewards, which is described as a tuple $\langle\statespace{},\agentspace{},\jointobservationspace_{\joint},\jointactionspace_{\joint},\statetransitionfunction{},\observationfunction{},\rewardfunction{},\discountfactor{}\rangle$ and is defined as follows:\\

\begin{itemize}
\setlength\itemsep{0em}

\item $\statespace{}$ is the finite set of global states in the environment, where $\state{}'\in \statespace{}$ denotes the next state of the current state $\state{}\in \statespace{}$. The state information is optionally available during training, but not available to the agents during execution.

\item $\agentspace{}=\{1,...,\numberofagents{}\}$ is the set of $\numberofagents{}$ agents. We use $\agentcounter{}\in\agentspace{}$ to denote the index of the agent.

\item $\jointobservationspace_{\joint}=\Pi_{\agentcounter{}\in\agentspace{}}\jointobservationspace{}_{\agentcounter{}}$ is the set of joint observations. At each timestep, a joint observation $\jointobservation{}=\langle\observation{}_1,...\observation{}_\numberofagents{}\rangle\in\jointobservationspace_{\joint}$ is received. Each agent $\agentcounter{}$ is only able to observe its individual observation $\observation{}_{\agentcounter{}}\in\jointobservationspace{}_{\agentcounter{}}$.

\item $\jointobservationhistoryspace_{\joint}=\Pi_{\agentcounter{}\in\agentspace{}}\jointobservationhistoryspace{}_{\agentcounter{}}$ is the set of joint action-observation histories. The joint history $\jointobservationhistory{}=\langle\observationhistory{}_1,...\observationhistory{}_\numberofagents{}\rangle\in\jointobservationhistoryspace_{\joint}$ concatenates all received observations and performed actions before a certain timestep, where $\observationhistory{}_{\agentcounter{}}\in\jointobservationhistoryspace{}_{\agentcounter{}}$ represents the action-observation history from agent $\agentcounter{}$.

\item $\jointactionspace_{\joint}=\Pi_{\agentcounter{}\in\agentspace{}}\jointactionspace{}_{\agentcounter{}}$ is the set of joint actions. At each timestep, the entire group of the agents take a joint action $\jointaction{}$, where $\jointaction{}=\langle\action{}_1,...,\action{}_{\numberofagents{}}\rangle\in\jointactionspace_{\joint}$. The individual action $\action{}_{\agentcounter{}}\in\jointactionspace_{\agentcounter{}}$ of each agent $\agentcounter{}$ is determined based on its stochastic policy $\policy{}_{\agentcounter{}}(\action{}_{\agentcounter{}}|\observationhistory{}_{\agentcounter{}}): \jointobservationhistoryspace{}_{\agentcounter{}}\times \jointactionspace{}_{\agentcounter{}}\rightarrow[0,1]$, expressed as $\action{}_{\agentcounter{}}\sim\policy{}_{\agentcounter{}}(\cdot|\observationhistory{}_{\agentcounter{}})$. Similarly, in single agent scenarios, we use $\action$ and $\action'$ to denote the actions of the agent at state $\state$ and $\state'$ under policy $\policy$, respectively.

\item $\timespace{}=\{1,...,\horizon{}\}$ represents the set of timesteps with horizon $\horizon{}$, where the index of the current timestep is denoted as $\timestep{}\in\timespace{}$. $\state^{\timestep}$, $\observation^{\timestep}$, $\observationhistory^{\timestep}$, and $\action^{\timestep}$ correspond to the environment information at timestep $\timestep$.

\item The transition function $\statetransitionfunction{}(\state{}'|\state{},\jointaction{}):\statespace{}\times\jointactionspace_{\joint}\times \statespace{}\rightarrow[0,1]$ specifies the state transition probabilities. Given $\state$ and $\jointaction$, the next state is denoted as $\state{}'\sim\statetransitionfunction{}(\cdot|\state{},\jointaction{})$.

\item The observation function $\observationfunction{}({\jointobservation{}}|\state{}):\jointobservationspace_{\joint}\times\statespace{}\rightarrow[0,1]$ specifies the joint observation probabilities. Given $\state{}$, the joint observation is represented as $\jointobservation{}\sim\observationfunction{}(\cdot|\state{})$.

\item $\rewardfunction{}(\reward|\state{},\jointaction{}):\statespace{}\times\jointactionspace_{\joint}\times\mathbb{R}\rightarrow[0,1]$ is the joint reward function shared among all agents. Given $\state$, the team reward is expressed as $\reward\sim\rewardfunction{}(\cdot|\state{},\jointaction{})$.

\item $\discountfactor{}\in\mathbb{R}$ is the discount factor with value within $(0, 1]$.
\end{itemize}
Under such an MARL formulation, this work concentrates on CTDE value function factorization methods, where the agents are trained in a centralized fashion and executed in a decentralized manner. In other words, the joint observation history $\jointobservationhistory{}$ is available during the learning processes of individual policies $[\policy{}_{\agentcounter{}}]_{\agentcounter{}\in\agentspace{}}$. During execution, each agent's policy $\policy{}_{\agentcounter{}}$ only conditions on its observation history $\observationhistory{}_{\agentcounter{}}$.


\subsection{IGM and Factorizable Tasks}
\label{subsec:background_igm_and_factorizable_task}

IGM is necessary for value function factorization~\cite{Son2019QTRAN}. For a joint action-value function $\utilityexp{}_{\joint{}}(\jointobservationhistory{},\jointaction{}): \jointobservationhistoryspace_{\joint} \times \jointactionspace_{\joint} \rightarrow\mathbb{R}$, if there exist $\numberofagents{}$ individual utility functions $[\utilityexp{}_\agentcounter{}(\observationhistory{}_\agentcounter{},\action{}_\agentcounter{}): \jointobservationhistoryspace{}_{\agentcounter{}} \times \jointactionspace{}_{\agentcounter{}} \rightarrow\mathbb{R}]_{\agentcounter{}\in\agentspace{}}$ such that the following condition holds:
\begin{equation}
\arg\max_\jointaction{} \utilityexp{}_{\joint{}}(\jointobservationhistory{},\jointaction{}) =
\begin{pmatrix}
\arg\max_{\action{}_1} \utilityexp{}_1(\observationhistory{}_1,\action{}_1)\\
\vdots \\
\arg\max_{\action{}_\numberofagents{}} \utilityexp{}_\numberofagents{}(\observationhistory{}_\numberofagents{},\action{}_\numberofagents{})
\end{pmatrix},
\label{eq:igm}
\end{equation}
then $[\utilityexp{}_\agentcounter{}]_{\agentcounter{}\in\agentspace{}}$ are said to satisfy IGM for $\utilityexp{}_{\joint{}}$ under $\jointobservationhistory{}$. In this case, we also say that $\utilityexp{}_{\joint{}}(\jointobservationhistory{},\jointaction{})$ is factorized by $[\utilityexp{}_\agentcounter{}(\observationhistory{}_\agentcounter{},\action{}_\agentcounter{})]_{\agentcounter{}\in\agentspace{}}$~\cite{Son2019QTRAN}. If $\utilityexp{}_{\joint{}}$ in a given task is factorizable under all $\jointobservationhistory{}\in \jointobservationhistoryspace_{\joint}$, we say that the task is factorizable. Intuitively, factorizable tasks indicate that there exists a factorization such that each agent can select the greedy action according to their individual utilities $[\utilityexp{}_\agentcounter{}]_{\agentcounter{}\in\agentspace{}}$ independently in a decentralized fashion. This enables the optimal individual actions to implicitly achieve the optimal joint action across the $\numberofagents{}$ agents. Since there is no individual reward, the factorized utilities do not estimate expected returns on their own \cite{Guestrin2001Utility} and are different from the value function definition commonly used in SARL.


\subsection{Value Function Factorization Methods}
\label{subsec:background_value_factorization_methods}
Based on IGM, value function factorization methods enable centralized training for factorizable tasks, while maintaining the ability for decentralized execution. In this work, we consider two such methods, VDN and QMIX, which can solve a subset of factorizable tasks that satisfies \additivity{} (Eq.~(\ref{eq:additivity})) and \monotonicity{} (Eq.~(\ref{eq:monotonicity})), respectively, given by:
\begin{equation} \utilityexp{}_{\joint{}}(\jointobservationhistory{},\jointaction{}) = \sum^{\numberofagents{}}_{\agentcounter{}=1} \utilityexp{}_\agentcounter{}(\observationhistory{}_\agentcounter{},\action{}_\agentcounter{}),
\label{eq:additivity}
\end{equation}
\begin{equation} \utilityexp{}_{\joint{}}(\jointobservationhistory{},\jointaction{}) = \monotonicfunction(\utilityexp{}_1(\observationhistory{}_1,\action{}_1),...,\utilityexp{}_\numberofagents{}(\observationhistory{}_\numberofagents{},\action{}_\numberofagents{})\vert\state),
\label{eq:monotonicity}
\end{equation}
where $\monotonicfunction$ is a monotonic function that satisfies $\frac{\partial \monotonicfunction}{\partial \utilityexp{}_\agentcounter{}}\ge 0, \forall \agentcounter{}\in\agentspace{}$, and conditions on the state $\state$ if the information is available during training. Either of these two equation is a sufficient condition for IGM~\cite{Son2019QTRAN}.


\subsection{Distributional RL}
\label{subsec:background_distributional_rl}

For notational simplicity, we consider a degenerated case with only a single agent, and the environment is fully observable until the end of Section~\ref{subsec:background_implicit_quantile_network}. Distributional RL generalizes classic expected RL methods by capturing the full return distribution $\utility{}(\state{},\action{})$ instead of the expected return $\utilityexp{}(\state{},\action{})$, and outperforms expected RL methods in various single-agent RL domains~\cite{Bellemare2017C51,Bellemare2019S51,Dabney2018QR-DQN,Dabney2018IQN,Rowland2019ER-DQN,Yang2019FQF}. Moreover, distributional RL enables improvements~\cite{Nikolov2019IDS,Zhang2019QUOTA,Mavrin2019DLTV} that require the information of the full return distribution. We define the distributional Bellman operator $\mathcal{T}^\pi$ as follows:
\begin{equation}
\mathcal{T}^\pi \utility{}(\state{},\action{})\stackrel{D}{:=}\rewardfunction{}(\state{},\action{})+\gamma \utility{}(\state{}',\action{}'),
\end{equation}
and the distributional Bellman optimality operator $\mathcal{T}^*$ as:
\begin{equation}
\mathcal{T}^* \utility{}(\state{},\action{})\stackrel{D}{:=}\rewardfunction{}(\state{},\action{})+\gamma \utility{}(\state{}',\action{}'^*),
\end{equation}
where $\action{}'^*=\arg\max_{\action'}\mathbb{E}[\utility{}(\state{}',\action')]$ is the optimal action at state $\state{}'$, and the expression $X\stackrel{D}{=}Y$ denotes that random variable $X$ and $Y$ follow the same distribution. Given some initial distribution $Z_0$, $\utility{}$ converges to the return distribution $\utility{}^\pi$ under $\pi$, contracting in terms of $p$-Wasserstein distance for all $p\in[1,\infty)$ by applying $\mathcal{T}^\pi$ repeatedly; while $\utility{}$ alternates between the optimal return distributions in the set $\mathcal{\utility{}}^*:=\{\utility^{\pi^*}:\pi^*\in\Pi^*\}$, under the set of optimal policies $\Pi^*$ by repeatedly applying $\mathcal{T}^*$~\cite{Bellemare2017C51}. The $p$-Wasserstein distance $W_p$ between the probability distributions of random variables $X$, $Y$ is given by:
\begin{equation}
W_p(X,Y)=\left(\int_0^1|\inversecdf_X(\quantile)-\inversecdf_Y(\quantile)|^p \mathrm{d}\quantile\right)^{1/p},
\end{equation}
where $(\inversecdf_X,\inversecdf_Y)$ are quantile functions of $(X,Y)$.


\subsection{Quantile Function and Quantile Regression}
\label{subsec:background_quantile_function_and_quantile_regression}

The relationship between the cumulative distribution function (CDF) $F_X$ and the quantile function $\quantilefunction_X$ (the generalized inverse CDF) of random variable $X$ is formulated as:
\begin{equation}
\quantilefunction_X(\quantile{})=\inf\{x\in\mathbb{R}:\quantile{}\le \cdf{}_X(x)\}, \forall\quantile\in[0,1].
\label{eq:quantile_function_to_inverse_cdf}
\end{equation}
The expectation of $X$ expressed in terms of $\quantilefunction_X(\quantile{})$ is:
\begin{equation}
\mathbb{E}[X]=\int_0^1\quantilefunction_X(\quantile{})\ \mathrm{d}\quantile.
\label{eq:expectation_of_quantile_function}
\end{equation}
In~\cite{Dabney2018QR-DQN}, the authors model the value function as a quantile function $\quantilefunction{}(\state{},\action{}\vert\quantile{})$. During optimization, a pair-wise sampled temporal difference (TD) error $\tderror$ for two quantile samples $\quantile{}, \quantile{}'\sim U([0,1])$ is defined as:
\begin{equation}
\tderror_{\timestep{}}^{\quantile{},\quantile{}'}=\reward{}+\gamma \quantilefunction{}(\state{}',\action{}'\vert\quantile{}') - \quantilefunction{}(\state{},\action{}\vert\quantile{}).
\end{equation}
The pair-wise loss $\asymmetrichuberloss{}^\kappa_{\quantile{}}$ is then defined based on the Huber quantile regression loss $\huberloss{}$~\cite{Dabney2018QR-DQN} with threshold $\kappa=1$, and is formulated as follows:
\begin{equation}
\asymmetrichuberloss{}^\kappa_{\quantile{}}(\tderror^{\quantile{},\quantile{}'})=|\quantile{}-\mathbb{I}\{\tderror^{\quantile{},\quantile{}'}<0\}|\frac{\huberloss{}(\tderror^{\quantile{},\quantile{}'})}{\kappa} \text{, with}
\end{equation}
\begin{equation}
\huberloss{}(\tderror^{\quantile{},\quantile{}'})=
\begin{cases}
    \frac{1}{2}(\tderror^{\quantile{},\quantile{}'})^2, & \text{if }|\tderror^{\quantile{},\quantile{}'}|\le\kappa\\
    \kappa(|\tderror^{\quantile{},\quantile{}'}|-\frac{1}{2}\kappa), & \text{otherwise}
\end{cases}.
\end{equation}
Given $\numberofquantiles{}$ quantile samples $[\quantile{}_i]^\numberofquantiles_{i=1}$ to be optimized with regard to $\numberofquantilesamples$  target quantile samples $[\quantile{}_j]^{\numberofquantilesamples}_{j=1}$, the total loss $\loss{}(\state{},\action{},\reward{},\state{}')$  is defined as the sum of the pair-wise losses, and is expressed as the following:
\begin{equation}
\loss{}(\state{},\action{},\reward{},\state{}')=\frac{1}{\numberofquantilesamples{}}\sum^{\numberofquantiles{}}_{i=1}\sum^{\numberofquantilesamples{}}_{j=1}\asymmetrichuberloss{}^\kappa_{\quantile{}_i}(\tderror^{\quantile{}_i,\quantile{}_j'}).
\label{eq:total_qr_loss}
\end{equation}


\subsection{Implicit Quantile Network}
\label{subsec:background_implicit_quantile_network}

Implicit quantile network (IQN)~\cite{Dabney2018IQN} is relatively light-weight when it is compared to other distributional RL methods. It approximates the return distribution $\utility{}(\state{},\action{})$ by an implicit quantile function $\quantilefunction{}(\state{},\action{}\vert\quantile{})=\implicitquantilefunction{}(\stateembeddingfunction{}(\state{}),\cosineembeddingfunction{}(\quantile{}))_\action{}$ for some differentiable functions $\implicitquantilefunction{}$, $\psi$, and $\phi$. Such an architecture is a type of universal value function approximator (UVFA)~\cite{Schaul2015UVFA}, which generalizes its predictions across states $\state{}\in\statespace{}$ and goals $\quantile{}\in [0,1]$, with the goals defined as different quantiles of the return distribution.
In practice, $\cosineembeddingfunction{}$ first expands the scalar $\quantile$ to an $n$-dimensional vector by $[\cos(\pi i\quantile)]^{\cosinedimension{}-1}_{i=0}$, followed by a single hidden layer with weights $[w_{ij}]$ and biases $[b_j]$ to produce a quantile embedding $\cosineembeddingfunction{}(\quantile{})=[\cosineembeddingfunction(\quantile{})_j]^{\text{dim}(\cosineembeddingfunction{}(\quantile{}))-1}_{j=0}$. The expression of $\cosineembeddingfunction(\quantile{})_j$ can be represented as the following:
\begin{equation}
\cosineembeddingfunction(\quantile{})_j:=\text{ReLU}(\sum_{i=0}^{\cosinedimension{}-1}\cos(\pi i\quantile)w_{ij}+b_j),
\end{equation}
where $\cosinedimension{}=64$ and $\text{dim}(\cosineembeddingfunction(\quantile{}))=\text{dim}(\stateembeddingfunction(\state{}))$. Then, $\cosineembeddingfunction{}(\quantile{})$ is combined with the state embedding $\stateembeddingfunction{}(\state{})$ by the element-wise (Hadamard) product ($\odot$), expressed as $\implicitquantilefunction:=\stateembeddingfunction{}\odot\cosineembeddingfunction{}$. The loss of IQN is defined as Eq.~(\ref{eq:total_qr_loss}) by sampling a batch of $\numberofquantiles{}$ and $\numberofquantilesamples{}$ quantiles from the policy network and the target network respectively.
During execution, the action with the largest expected return $\utilityexp{}(\state{},\action{})$ is chosen. Since IQN does not model the expected return explicitly, $\utilityexp{}(\state{},\action{})$ is approximated by calculating the mean of the sampled return through $\numberofquantiletestsamples{}$ quantile samples $\hat\quantile{}_i\sim U([0,1]), \forall i\in[1,\numberofquantiletestsamples{}]$ based on Eq.~(\ref{eq:expectation_of_quantile_function}), expressed as follows:
\begin{equation}
\utilityexp{}(\state{},\action{})=\int_0^1\quantilefunction{}(\state{},\action{}\vert\quantile{})\ \mathrm{d}\quantile\approx\frac{1}{\numberofquantiletestsamples{}}\sum_{i=1}^{\numberofquantiletestsamples{}}\quantilefunction{}(\state{},\action{}\vert\hat\quantile{}_i).
\end{equation}


\subsection{Quantile Mixture}
\label{subsec:background_quantile_mixture}

Multiple quantile functions (e.g., IQNs) sharing the same quantile $\quantile{}$ may be combined into a single quantile function $\quantilefunction{}(\quantile{})$, in a form of quantile mixture expressed as follows:
\begin{equation}
\quantilefunction{}(\quantile{})=\sum^{\numberofagents{}}_{k=1}\modelparameter{}_{\agentcounter{}} \quantilefunction_{\agentcounter{}}(\quantile{}),
\label{eq:quantile_mixture}
\end{equation}
where $[\quantilefunction_{\agentcounter{}}(\quantile{})]_{\agentcounter{}\in\agentspace{}}$ are quantile functions, and $[\modelparameter{}_{\agentcounter{}}]_{\agentcounter{}\in\agentspace{}}$ are model parameters~\cite{Karvanen2006QuantileMixture}. The condition for  $[\modelparameter{}_{\agentcounter{}}]_{\agentcounter{}\in\agentspace{}}$ is that $\quantilefunction{}(\quantile{})$ must satisfy the properties of a quantile function. The concept of quantile mixture is analogous to the mixture of multiple probability density functions (PDFs), expressed as follows:
\begin{equation}
\pdf{}(x)=\sum^{\numberofagents{}}_{\agentcounter{}=1}\pdfmodelparameter{}_{\agentcounter{}} \pdf{}_{\agentcounter{}}(x),
\end{equation}
where $[\pdf{}_{\agentcounter{}}(x)]_{\agentcounter{}\in\agentspace{}}$ are PDFs, $\sum^{\numberofagents{}}_{\agentcounter{}=1}\pdfmodelparameter{}_{\agentcounter{}}=1$, and $\pdfmodelparameter{}_{\agentcounter{}}\ge 0$.

\section{Methodology}
\label{sec:methodology}

In this section, we walk through the proposed DFAC framework and its derivation procedure. We first discuss a naive distributional factorization and its limitation in Section~\ref{subsec:methodology_distributional_igm_condition}. Then, we introduce the DFAC framework to address the limitation, and show that DFAC is able to generalize distributional RL to all factorizable tasks in Section~\ref{subsec:methodology_the_proposed_dfac_framework}. After that, \ddn{} and \dmix{} are introduced as the DFAC variants of VDN and QMIX, respectively, in Section~\ref{subsec:methodology_distributional_variant_of_vdn_and_qmix}. Finally, a practical implementation of DFAC based on quantile mixture is presented in Section~\ref{subsec:methodology_a_practical_implementation_of_dfac}. All proofs of the theorems in this section are provided in the supplementary material.


\subsection{Distributional IGM}
\label{subsec:methodology_distributional_igm_condition}

Since IGM is necessary for value function factorization, a distributional factorization that satisfies IGM is required for factorizing stochastic value functions. We first discuss a naive distributional factorization that simply replaces deterministic utilities $\utilityexp$ with stochastic utilities $\utility$. Then, we provide a theorem to show that the naive distributional factorization is insufficient to guarantee the IGM condition.

\begin{definition}[Distributional IGM]
\label{def:distributional_igm}

A finite number of individual stochastic utilities $[\utility_{\agentcounter}(\observationhistory_{\agentcounter}, \action_{\agentcounter})]_{\agentcounter \in \agentspace}$, are said to satisfy Distributional IGM (\digm{}) for a stochastic joint action-value function $\utility_{\joint}(\jointobservationhistory, \jointaction{})$ under $\jointobservationhistory$, if $[\mathbb{E}[\utility_{\agentcounter}(\observationhistory_{\agentcounter}, u_{\agentcounter})]]_{\agentcounter \in \agentspace}$ satisfy IGM for $\mathbb{E}[\utility_{\joint}(\jointobservationhistory, \jointaction{})]$ under $\jointobservationhistory$, represented as:
\begin{equation}
\arg\max_\jointaction{} \mathbb{E}[\utility{}_{\joint{}}(\jointobservationhistory{},\jointaction{})] =
\begin{pmatrix}
\arg\max_{\action{}_1} \mathbb{E}[\utility{}_1(\observationhistory{}_1,\action{}_1)]\\
\vdots \\
\arg\max_{\action{}_\numberofagents{}} \mathbb{E}[\utility{}_\numberofagents{}(\observationhistory{}_\numberofagents{},\action{}_\numberofagents{})]
\end{pmatrix}.
\notag{}
\end{equation}
\end{definition}

\begin{theorem}
\label{thm:distributional_igm}

Given a deterministic joint action-value function $\utilityexp_{\joint}$, a stochastic joint action-value function $\utility_{\joint}$, and a factorization function $\meandecompositionfunction$ for deterministic utilities:
\begin{equation}
\utilityexp{}_{\joint{}}(\jointobservationhistory{},{\jointaction{}})=\meandecompositionfunction(\utilityexp{}_1(\observationhistory{}_1,\action{}_1), ..., \utilityexp{}_{\numberofagents{}}(\observationhistory{}_{\numberofagents{}},\action{}_{\numberofagents{}})\vert\state),
\notag{}
\end{equation}
such that $[\utilityexp{}_\agentcounter{}]_{\agentcounter{}\in\agentspace{}}$ satisfy IGM for $\utilityexp{}_{\joint{}}$ under $\jointobservationhistory{}$, the following distributional factorization:
\begin{equation}
\utility{}_{\joint{}}(\jointobservationhistory{},{\jointaction{}})=\meandecompositionfunction(\utility{}_1(\observationhistory{}_1,\action{}_1), ..., \utility{}_{\numberofagents{}}(\observationhistory{}_{\numberofagents{}},\action{}_{\numberofagents{}})\vert\state).
\notag{}
\end{equation}

is insufficient to guarantee that $[\utility{}_\agentcounter{}]_{\agentcounter{}\in\agentspace{}}$ satisfy \digm{} for $\utility{}_{\joint{}}$ under $\jointobservationhistory{}$.

\end{theorem}

In order to satisfy \digm{} for stochastic utilities, an alternative factorization strategy is necessary.


\subsection{The Proposed DFAC Framework}
\label{subsec:methodology_the_proposed_dfac_framework}

We propose Mean-Shape Decomposition and the DFAC framework to ensure that \digm{} is satisfied for stochastic utilities.

\begin{definition}[Mean-Shape Decomposition]
\label{def:mean_shape_decomposition}

A given random variable $Z$ can be decomposed as follows:
\begin{equation}
\begin{split}
Z &= \mathbb{E}[Z]+(Z-\mathbb{E}[Z]) \\
&= Z_{\mathrm{mean}}+Z_{\mathrm{shape}}\ ,
\end{split}
\notag{}
\end{equation}
where $\mathrm{Var}(Z_{\mathrm{mean}})=0$ and $\mathbb{E}[Z_{\mathrm{shape}}]=0$.

\end{definition}

We propose DFAC to decompose a joint return distribution $\utility{}_{\joint{}}$ into its deterministic part $\utility{}_{\text{mean}}$ (i.e., expected value) and stochastic part $\utility{}_{\text{shape}}$ (i.e., higher moments), which are approximated by two different functions $\meandecompositionfunction$ and $\shapedecompositionfunction$, respectively. The factorization function $\meandecompositionfunction$ is required to precisely factorize the expectation of $\utility{}_{\joint{}}$ in order to satisfy \digm{}. On the other hand, the shape function $\shapedecompositionfunction$ is allowed to roughly factorize the shape of $\utility{}_{\joint{}}$, since the main objective of modeling the return distribution is to assist non-linear approximation of the expectation of $\utility{}_{\joint{}}$ \cite{Lyle2019Comparative}, rather than accurately model the shape of $\utility{}_{\joint{}}$.

\begin{theorem}[DFAC Theorem]
\label{thm:dfac}

Given a deterministic joint action-value function $\utilityexp_{\joint}$, a stochastic joint action-value function $\utility_{\joint}$, and a factorization function $\meandecompositionfunction$ for deterministic utilities:
\begin{equation}
\utilityexp{}_{\joint{}}(\jointobservationhistory{},{\jointaction{}})=\meandecompositionfunction(\utilityexp{}_1(\observationhistory{}_1,\action{}_1), ..., \utilityexp{}_{\numberofagents{}}(\observationhistory{}_{\numberofagents{}},\action{}_{\numberofagents{}})\vert\state),
\notag{}
\end{equation}
such that $[\utilityexp{}_\agentcounter{}]_{\agentcounter{}\in\agentspace{}}$ satisfy IGM for $\utilityexp{}_{\joint{}}$ under $\jointobservationhistory{}$, by Mean-Shape Decomposition, the following distributional factorization:
\begin{equation}
\begin{split}
\utility{}_{\joint{}}(\jointobservationhistory{},{\jointaction{}}) &= \mathbb{E}[\utility{}_{\joint{}}(\jointobservationhistory{},{\jointaction{}})]+(\utility{}_{\joint{}}(\jointobservationhistory{},{\jointaction{}})-\mathbb{E}[\utility{}_{\joint{}}(\jointobservationhistory{},{\jointaction{}})]) \\
&= \utility{}_{\mathrm{mean}}(\jointobservationhistory{},{\jointaction{}})+\utility{}_{\mathrm{shape}}(\jointobservationhistory{},{\jointaction{}})\\
&= \meandecompositionfunction{}(\utilityexp_1(\observationhistory{}_1,\action{}_1),...,\utilityexp_{\numberofagents{}}(\observationhistory{}_{\numberofagents{}},\action{}_{\numberofagents{}})\vert\state)\\
&+ \shapedecompositionfunction{}(\utility_1(\observationhistory{}_1,\action{}_1),...,\utility_{\numberofagents{}}(\observationhistory{}_{\numberofagents{}},\action{}_{\numberofagents{}})\vert\state).
\end{split}
\notag{}
\end{equation}
is sufficient to guarantee that $[\utility{}_\agentcounter{}]_{\agentcounter{}\in\agentspace{}}$ satisfy \digm{} for $\utility{}_{\joint{}}$ under $\jointobservationhistory{}$, where $\mathrm{Var}(\meandecompositionfunction)=0$ and  $\mathbb{E}[\shapedecompositionfunction]=0$.
\end{theorem}

Theorem.~\ref{thm:dfac} reveals that the choice of $\meandecompositionfunction$ determines whether IGM holds, regardless of the choice of $\shapedecompositionfunction$, as long as $\mathbb{E}[\shapedecompositionfunction]=0$. Under this setting, any differentiable factorization function of deterministic variables can be extended to a factorization function of random variables. Such a decomposition enables approximation of joint distributions for all factorizable tasks under appropriate choices of  $\meandecompositionfunction{}$ and $\shapedecompositionfunction{}$.


\begin{figure*}[t]
\includegraphics[width=\linewidth]{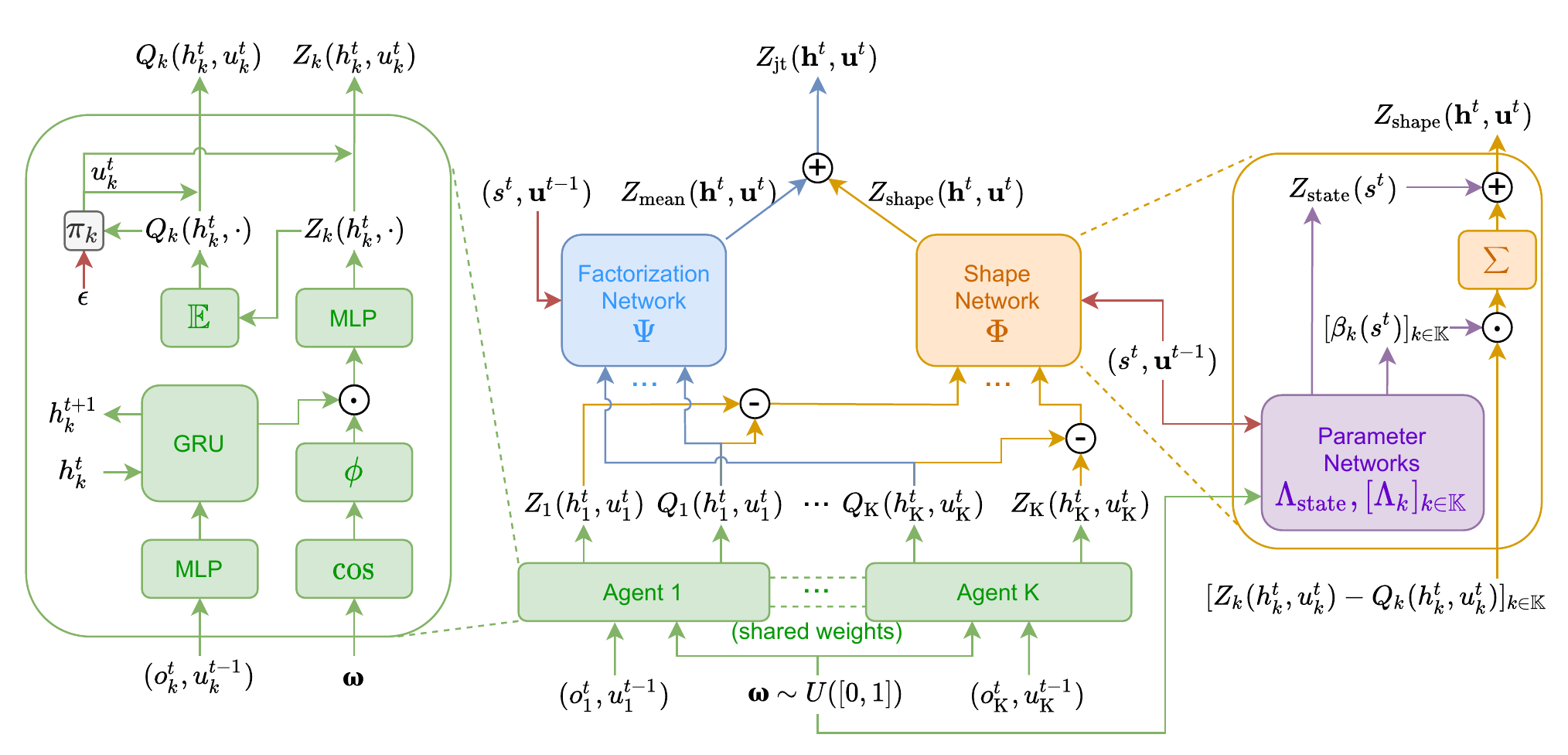}
\caption{The DFAC framework consists of a factorization network $\meandecompositionfunction{}$ and a shape network $\shapedecompositionfunction$ for decomposing the deterministic part $\utility{}_{\text{mean}}$ (i.e., $\utilityexp{}_{\joint{}}$) and the stochastic part $\utility{}_{\text{shape}}$ of the total return distribution $\utility{}_{\joint{}}$, as described in Theorem~\ref{thm:dfac}. The shape network contains parameter networks $\additionalparameterfunction_{\text{state}}(\state{};\quantile{})$ and $[\additionalparameterfunction_{\agentcounter}(\state{})]_{\agentcounter{}\in\agentspace{}}$ for generating  $\utility_{\text{state}}(\state{})$ and $\modelparameter{}_{\agentcounter}(s)$.}
\label{fig:dfac}
\end{figure*}

\subsection{A Practical Implementation of DFAC}
\label{subsec:methodology_a_practical_implementation_of_dfac}

In this section, we provide a practical implementation of the shape function $\shapedecompositionfunction$ in DFAC, effectively extending any differentiable factorization function $\meandecompositionfunction{}$ (e.g., the additive function of VDN, the monotonic mixing network of QMIX, etc.) that satisfies the IGM condition into its DFAC variant.

Theoretically, the sum of random variables appeared in \ddn{} and \dmix{} can be described precisely by a joint CDF. However, the exact derivation of this joint CDF is usually computationally expensive and impractical~\cite{Lin2019DR-DRL}. As a result, DFAC utilizes the property of quantile mixture to approximate the shape function $\shapedecompositionfunction{}$ in $\mathrm{O}(\numberofagents\numberofquantiles)$ time.

\begin{theorem}
\label{thm:sum_of_rv}
Given a quantile mixture:
\begin{equation}
\quantilefunction(\quantile{})=\sum^{\numberofagents{}}_{k=1}\modelparameter{}_{\agentcounter{}} \quantilefunction_{\agentcounter{}}(\quantile{})
\notag{}
\end{equation}
with $\numberofagents{}$ components $[\quantilefunction_{\agentcounter{}}]_{\agentcounter{}\in\agentspace{}}$ and non-negative model parameters $[\modelparameter{}_{\agentcounter{}}]_{\agentcounter{}\in\agentspace{}}$. There exist a set of random variables $\utility{}$ and $[\utility_{\agentcounter{}}]_{\agentcounter{}\in\agentspace{}}$  corresponding to the quantile functions $\quantilefunction$ and $[\quantilefunction_{\agentcounter{}}]_{\agentcounter{}\in\agentspace{}}$, respectively, with the following relationship:
\begin{equation}
\utility{}=\sum_{\agentcounter\in\agentspace{}}\modelparameter{}_{\agentcounter{}} \utility{}_{\agentcounter{}}.
\notag{}
\end{equation}
\end{theorem}

Based on Theorem~\ref{thm:sum_of_rv}, the quantile function $\quantilefunction_{\mathrm{shape}}$ of $\utility_{\mathrm{shape}}$ in DFAC can be approximated by the following:
\begin{equation}
\begin{split}
\quantilefunction_{\mathrm{shape}}(\jointobservationhistory{},{\jointaction{}}\vert\quantile{})=\ &\quantilefunction_{\text{state}}(\state\vert\quantile{})\\
+\sum_{\agentcounter\in\agentspace{}}\modelparameter{}_{\agentcounter}(s)(&\quantilefunction_{\agentcounter}(\observationhistory{}_{\agentcounter{}},\action{}_{\agentcounter{}}\vert\quantile{})-\utilityexp_{\agentcounter}(\observationhistory{}_{\agentcounter{}},\action{}_{\agentcounter{}})),
\label{eq::dfac_approximation_equation}
\end{split}
\end{equation}
where $\quantilefunction_{\text{state}}(\state{}\vert\quantile{})$ and $[\modelparameter{}_{\agentcounter}(s)]_{\agentcounter{}\in\agentspace{}}$ are respectively generated by function approximators $\additionalparameterfunction_{\text{state}}(\state{}\vert\quantile{})$ and $[\additionalparameterfunction_{\agentcounter}(\state{})]_{\agentcounter{}\in\agentspace{}}$, satisfying constraints $\modelparameter{}_{\agentcounter}(s)\ge0, \forall\agentcounter{\in\agentspace{}}$ and $\int_0^1\quantilefunction_{\text{state}}(\state\vert\quantile{})\ \mathrm{d}\quantile=0$. The term $\quantilefunction_{\text{state}}$ models the shape of an additional state-dependent utility (introduced by QMIX at the last layer of the mixing network), which extends the state-dependent bias in QMIX to a full distribution. The full network architecture of DFAC is illustrated in Fig.~\ref{fig:dfac}.

This transformation enables DFAC to decompose the quantile representation of a joint distribution into the quantile representations of individual utilities. In this work, $\shapedecompositionfunction{}$ is implemented by a large IQN composed of multiple IQNs, optimized through the loss function defined in Eq.~(\ref{eq:total_qr_loss}).

\subsection{DFAC Variant of VDN and QMIX}
\label{subsec:methodology_distributional_variant_of_vdn_and_qmix}

In order to validate the proposed DFAC framework, we next discuss the DFAC variants of two representative factorization methods: VDN and QMIX. \ddn{} extends VDN to its DFAC variant, expressed as:
\begin{equation}
\begin{split}
\utility_\joint=\sum_{\agentcounter\in\agentspace{}}\utilityexp_{\agentcounter}+\sum_{\agentcounter\in\agentspace{}}(\utility_{\agentcounter}-\utilityexp_{\agentcounter}), \text{ given}
\end{split}
\end{equation}
$\utility_{\mathrm{mean}}=\sum_{\agentcounter\in\agentspace{}}\utilityexp_{\agentcounter}$, $\utility_{\mathrm{shape}}=\sum_{\agentcounter\in\agentspace{}}(\utility_{\agentcounter}-\utilityexp_{\agentcounter})$; while \dmix{} extends QMIX to its DFAC variant, expressed as:
\begin{equation}
\utility_\joint=\monotonicfunction{}(\utilityexp_1,...,\utilityexp_\numberofagents\vert\state)+\sum_{\agentcounter\in\agentspace{}}(\utility_{\agentcounter}-\utilityexp_{\agentcounter}), \text{ given}
\end{equation}
$\utility_{\mathrm{mean}}=\monotonicfunction{}(\utilityexp_1,...,\utilityexp_\numberofagents\vert\state)$, $\utility_{\mathrm{shape}}=\sum_{\agentcounter\in\agentspace{}}(\utility_{\agentcounter}-\utilityexp_{\agentcounter})$.

Both \ddn{} and \dmix{} choose $\quantilefunction_{\text{state}}=0$ and $[\modelparameter{}_{\agentcounter}=1]_{\agentcounter{}\in\agentspace{}}$ for simplicity. Automatically learning the values of $\quantilefunction_{\text{state}}$ and $[\modelparameter{}_{\agentcounter}]_{\agentcounter{}\in\agentspace{}}$ is proposed as future work.
\section{A Stochastic Two-Step Game}
\label{sec:distributional_2_step_game}

\begin{figure*}[t]
\begin{tabular}{cc}
\begin{minipage}{0.47\textwidth}
\scriptsize
\setlength{\extrarowheight}{3pt}
\centering
\captionof{table}{
An illustration of the flow of the stochastic two-step game. Each agent is able to perform an action from $\{A,B\}$ in each step, with a subscript denoting the agent index. In the first step, action $A_1$ takes the agents from the initial \texttt{State 1} to \texttt{State 2A}, while action $B_1$ takes them to \texttt{State 2B} instead. The transitions from \texttt{State 1} to \texttt{State 2A} or \texttt{State 2B} yield zero reward. In the second step, the global rewards are sampled from the normal distributions defined in the payoff matrices.
}\vspace{0.5cm}
\begin{tikzpicture}[overlay]
    \node[] at (0,0) {};
    \node[] at (4.6,1.15) {\texttt{State $1$}};
    \draw[-latex] (4.6,1.15-0.2) -- (4.6-0.8,0.4) node[] [midway, above, sloped] {\cc{$A_1$}};
    \draw[-latex] (4.6,1.15-0.2) -- (4.6+0.8,0.4) node[] [midway, above, sloped] {\cc{$B_1$}};
\end{tikzpicture}
\begin{tabular}{cc|*{2}{>{\centering\arraybackslash}p{.11\linewidth}|}}
	& \multicolumn{1}{c}{} & \multicolumn{2}{c}{\bb{Agent $2$}} \\
	& \multicolumn{1}{c}{} & \multicolumn{1}{c}{\bb{$A_2$}}  & \multicolumn{1}{c}{\bb{$B_2$}} \\ 
	\cline{3-4}
    \multirow{2}{*}{\rotatebox[origin=c]{90}{\cc{Agent $1$}}} & \cc{$A_1$} & $\mathcal{N}(7,0)$ & $\mathcal{N}(7,0)$ \\ \cline{3-4}
    & \cc{$B_1$} & $\mathcal{N}(7,0)$ & $\mathcal{N}(7,0)$  \\\cline{3-4}
    & \multicolumn{1}{c}{}  & \multicolumn{2}{c}{\texttt{State $2$A}} \\
\end{tabular}
~
\begin{tabular}{c|*{2}{>{\centering\arraybackslash}p{.12\linewidth}|}}
\multicolumn{1}{c}{} & \multicolumn{2}{c}{\bb{Agent $2$}} \\
\multicolumn{1}{c}{} & \multicolumn{1}{c}{\bb{$A_2$}}  & \multicolumn{1}{c}{\bb{$B_2$}} \\ 
\cline{2-3}
\cc{$A_1$} & $\mathcal{N}(0,2)$ & $\mathcal{N}(1,13)$ \\ \cline{2-3}
\cc{$B_1$} & $\mathcal{N}(1,13)$ & $\mathcal{N}(8,29)$  \\ \cline{2-3}
\multicolumn{1}{c}{} & \multicolumn{2}{c}{\texttt{State $2$B}} \\
\end{tabular}
\label{table:2-step-game-dmix}
\end{minipage}
\begin{tabular}{cc}
\begin{minipage}{0.45\textwidth} 
\includegraphics[width=\linewidth]{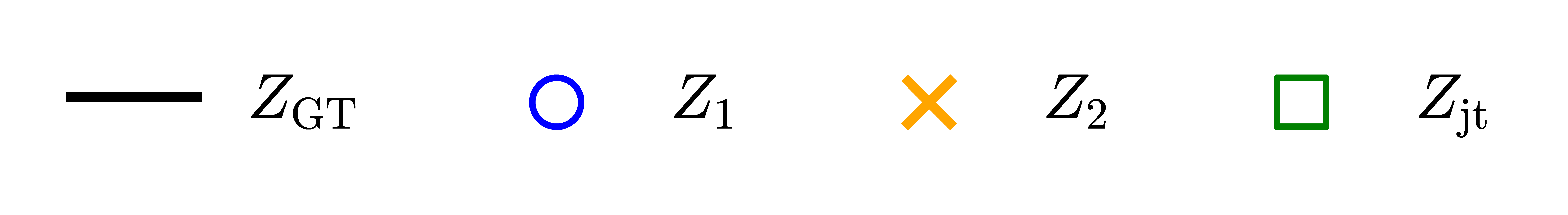}\vspace{-2em}
\label{fig:two_step_game_legend}
\end{minipage} \\
\begin{minipage}{0.47\textwidth}
\begin{subfigure}[t]{0.52\linewidth}
\includegraphics[width=4.25cm]{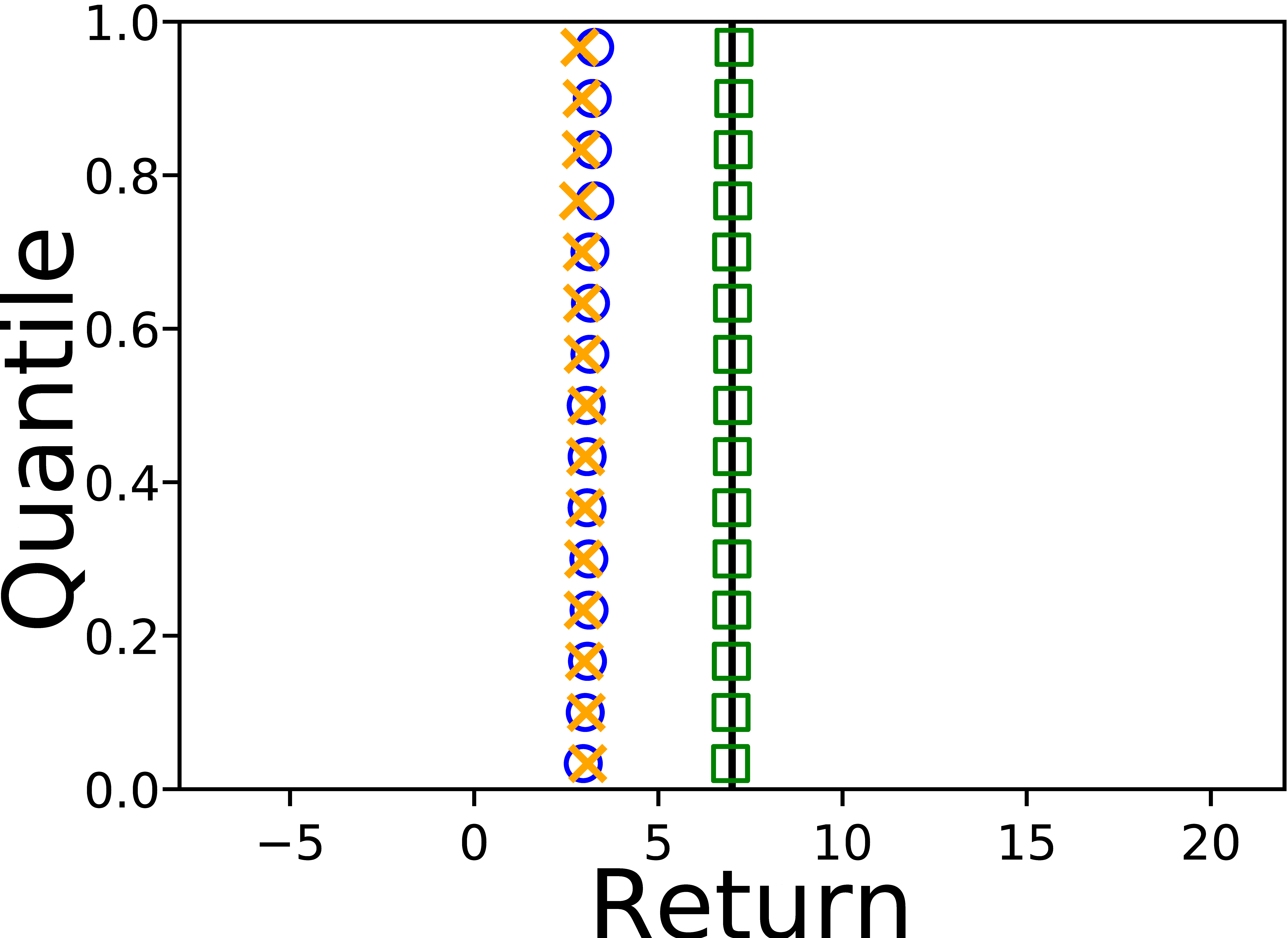}
\subcaption{$\langle B_1,B_2\rangle$ at \texttt{State $2$A}}
\end{subfigure}
\begin{subfigure}[t]{0.47\linewidth}
\includegraphics[width=3.74cm]{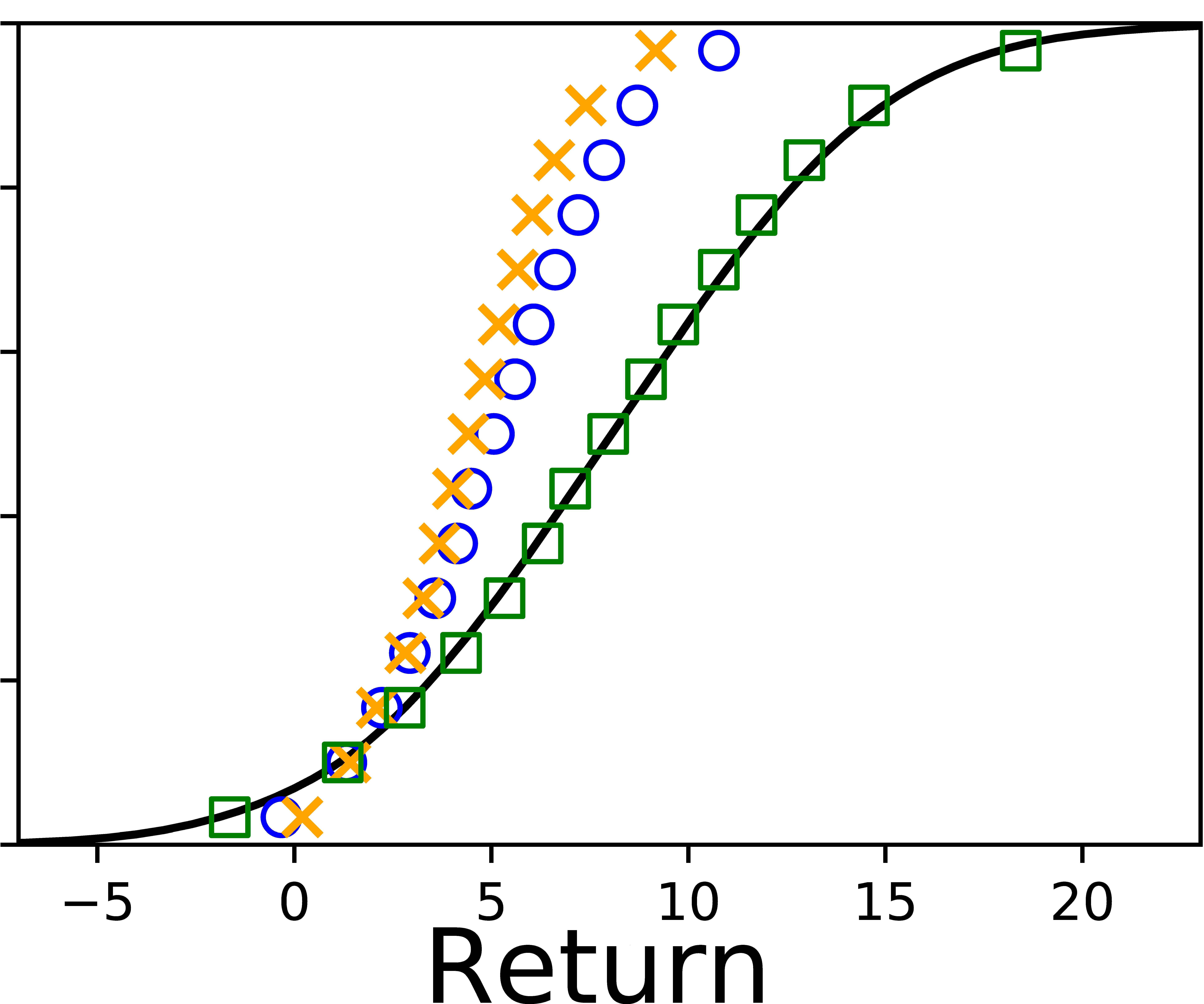}
\subcaption{$\langle B_1,B_2\rangle$ at \texttt{State $2$B}}
\end{subfigure}
\caption{
(a) and (b) plot the value function factorization of the joint action $\langle B_1,B_2\rangle$ in \texttt{State~2A} and \texttt{State~2B}. The black line/curve shows the true return CDFs. The blue circles and the orange cross marks depict agent 1’s and agent 2’s learned utility, respectively, while the green squares indicate the estimated joint return.
}
\label{fig:2-step-game-dmix-results}
\end{minipage}
\end{tabular}
\end{tabular}
\end{figure*}

In the previous expected value function factorization methods (e.g., VDN, QMIX, etc.), the factorization is achieved by modeling $\utilityexp{}_{\joint{}}$ and $[\utilityexp{}_{\agentcounter{}}]_{\agentcounter{}\in\agentspace{}}$ as deterministic variables, overlooking the information of higher moments in the full return distributions $\utility{}_{\joint{}}$ and $[\utility{}_{\agentcounter{}}]_{\agentcounter{}\in\agentspace{}}$.
In order to demonstrate DFAC's ability of factorization, we begin with a toy example modified from~\cite{Rashid2018QMIX} to show that DFAC is able to approximate the true return distributions, and factorize the mean and variance of the approximated total return $\utility{}_{\joint{}}$ into utilities $[\utility{}_{\agentcounter{}}]_{\agentcounter{}\in\agentspace{}}$.
Table~\ref{table:2-step-game-dmix} illustrates the flow of a two-step game consisting of two agents and three states \texttt{1}, \texttt{2A}, and \texttt{2B}, where \texttt{State~1} serves as the initial state, and each agent is able to perform an action from $\{A, B\}$ in each step. In the first step (i.e., \texttt{State 1}), the action of agent $1$ (i.e., actions $A_1$ or $B_1$) determines which of the two matrix games (\texttt{State 2A} or \texttt{State 2B}) to play in the next step, regardless of the action performed by agent $2$ (i.e., actions $A_2$ or $B_2$). For all joint actions performed in the first step, no reward is provided to the agents.
In the second step, both agents choose an action and receive a global reward according to the payoff matrices depicted in Table~\ref{table:2-step-game-dmix}, where the global rewards are sampled from a normal distribution $\mathcal{N}(\mu,\sigma^2)$ with mean $\mu$ and standard deviation $\sigma$. The hyperparameters of the two-step game are offered in the supplementary material in detail.

Table~\ref{table:2-step-game-dmix-results} presents the learned factorization of \dmix{} for each state after convergence, where the first rows and the first columns of the tables correspond to the factorized distributions of the individual utilities (i.e., $\utility_1$ and $\utility_2$), and the main content cells of them correspond to the joint return distributions (i.e., $\utility_{\joint}$). 
From Tables~\ref{table:2-step-game-dmix-results}(b) and~\ref{table:2-step-game-dmix-results}(c), it is observed that no matter the true returns are deterministic (i.e., \texttt{State 2A}) or stochastic (i.e., \texttt{State 2B}), \dmix{} is able to approximate the true returns in Table~\ref{table:2-step-game-dmix} properly, which are not achievable by expected value function factorization methods. The results demonstrate DFAC's ability to factorize the joint return distribution rather than expected returns. \dmix{}'s ability to reconstruct the optimal joint policy in the two-step game further shows that \dmix{} can represent the same set of tasks as QMIX.

To further illustrate DFAC's capability of factorization, Figs.~\ref{fig:2-step-game-dmix-results}(a)-\ref{fig:2-step-game-dmix-results}(b) visualize the factorization of the joint action $\langle B_1,B_2\rangle$ in \texttt{State $2$A} and $\langle B_1,B_2\rangle$ in \texttt{State $2$B}, respectively.
As IQN approximates the utilities $\utility{}_1$ and $\utility{}_2$ implicitly,  $\utility{}_1$, $\utility{}_2$, and $\utility{}_{\joint{}}$ can only be plotted in terms of samples. $\utility{}_{\joint{}}$ in Fig.~\ref{fig:2-step-game-dmix-results}(a) shows that \dmix{} degenerates to QMIX when approximating deterministic returns (i.e., $\mathcal{N}(7,0)$), while $\utility{}_{\joint{}}$ in Fig.~\ref{fig:2-step-game-dmix-results}(b) exhibits \dmix{}'s ability to capture the uncertainty in stochastic returns (i.e., $\mathcal{N}(8,29)$).

    \begin{table*}[t]
\caption{
The learned factorization of \dmix{}.
All of the cells show the sampled mean $\mu$ and the sampled variance
$\sigma^2$ with Bessel’s correction.
The main content cells correspond to the joint return distributions for different combinations of states and actions.
The first columns and first rows of these tables correspond to the distributions of the utilities for agents 1 and 2, respectively.
The top-left cells of these tables are the state-dependent utility $V$. DFAC enables the approximation of the true joint return distributions in Table~1, and allows them to be factorized into the distributions of the utilities for the agents.}
\footnotesize
\setlength{\extrarowheight}{10pt}
\centering
\begin{minipage}{0.36\textwidth}
\scriptsize
\begin{tabular}{c|c||c|c|}
\multicolumn{1}{c}{} & \multicolumn{1}{c}{\texttt{State $1$}} & \multicolumn{1}{c}{\bb{$A_2$}}  & \multicolumn{1}{c}{\bb{$B_2$}} \\ \cline{2-4}
\pp{$V$}  & $\begin{aligned} \pp{\mu}&=-0.32 \\ \pp{\sigma^2}&=0.00 \end{aligned}$ & $\begin{aligned} \bb{\mu}&=2.66 \\ \bb{\sigma^2}&=0.10 \end{aligned}$ & $\begin{aligned} \bb{\mu}&=2.65 \\ \bb{\sigma^2}&=0.10 \end{aligned}$ \\ \hhline{~===}
\cc{$A_1$} & $\begin{aligned} \cc{\mu}&=2.56 \\ \cc{\sigma^2}&=0.08 \end{aligned}$ & $\begin{aligned} \mu&=6.94 \\ \sigma^2&=0.00 \end{aligned}$ & $\begin{aligned} \mu&=6.92 \\ \sigma^2&=0.00 \end{aligned}$ \\ \cline{2-4}
\cc{$B_1$} & $\begin{aligned} \cc{\mu}&=3.58 \\ \cc{\sigma^2}&=19.11 \end{aligned}$ & $\begin{aligned} \mu&=7.94 \\ \sigma^2&=21.85 \end{aligned}$ & $\begin{aligned} \mu&=7.92 \\ \sigma^2&=21.86 \end{aligned}$ \\ \cline{2-4}
\end{tabular}
\ 
\subcaption{Learned utilities of \texttt{State 1}}
\end{minipage}
~
\setlength{\extrarowheight}{10pt}
\centering
\begin{minipage}{0.29\textwidth}
\scriptsize
\begin{tabular}{|c||c|c|}
\multicolumn{1}{c}{\texttt{State $2$A}} & \multicolumn{1}{c}{\bb{$A_2$}}  & \multicolumn{1}{c}{\bb{$B_2$}} \\ \cline{1-3}
$\begin{aligned} \pp{\mu}&=0.49 \\ \pp{\sigma^2}&=0.00 \end{aligned}$ & $\begin{aligned} \bb{\mu}&=1.76 \\ \bb{\sigma^2}&=0.00 \end{aligned}$ & $\begin{aligned} \bb{\mu^{}}&=1.75 \\ \bb{\sigma^2}&=0.00 \end{aligned}$ \\ \hhline{===}
$\begin{aligned} \cc{\mu}&=2.09 \\ \cc{\sigma^2}&=0.00 \end{aligned}$ & $\begin{aligned} \mu&=7.01 \\ \sigma^2&=0.00 \end{aligned}$ & $\begin{aligned} \mu&=6.99 \\ \sigma^2&=0.00 \end{aligned}$ \\ \cline{1-3}
$\begin{aligned} \cc{\mu}&=2.09 \\ \cc{\sigma^2}&=0.00 \end{aligned}$ & $\begin{aligned} \mu&=7.01 \\ \sigma^2&=0.00 \end{aligned}$ & $\begin{aligned} \mu&=6.99 \\ \sigma^2&=0.00 \end{aligned}$ \\ \cline{1-3}
\end{tabular}
\ 
\subcaption{Learned utilities of \texttt{State 2A}}
\end{minipage}
~
\setlength{\extrarowheight}{10pt}
\centering
\begin{minipage}{0.29\textwidth}
\scriptsize
\begin{tabular}{|c||c|c|}
\multicolumn{1}{c}{\texttt{State $2$B}} & \multicolumn{1}{c}{\bb{$A_2$}}  & \multicolumn{1}{c}{\bb{$B_2$}} \\ \cline{1-3}
$\begin{aligned} \pp{\mu}&=0.38 \\ \pp{\sigma^2}&=0.00 \end{aligned}$ & $\begin{aligned} \bb{\mu}&=-4.55 \\ \bb{\sigma^2}&=0.29 \end{aligned}$ & $\begin{aligned} \bb{\mu^{}}&=3.08 \\ \bb{\sigma^2}&=5.87 \end{aligned}$ \\ \hhline{===}
$\begin{aligned} \cc{\mu}&=-3.50 \\ \cc{\sigma^2}&=0.40 \end{aligned}$ & $\begin{aligned} \mu&=-0.05 \\ \sigma^2&=1.37 \end{aligned}$ & $\begin{aligned} \mu&=1.01 \\ \sigma^2&=9.30 \end{aligned}$ \\ \cline{1-3}
$\begin{aligned} \cc{\mu}&=3.52 \\ \cc{\sigma^2}&=6.81 \end{aligned}$ & $\begin{aligned} \mu&=1.24 \\ \sigma^2&=9.86 \end{aligned}$ & $\begin{aligned} \mu&=8.14 \\ \sigma^2&=25.30 \end{aligned}$ \\ \cline{1-3}
\end{tabular}
\ 
\subcaption{Learned utilities of \texttt{State 2B}}
\end{minipage}
\label{table:2-step-game-dmix-results}
\end{table*}
\section{Experiment Results on SMAC}
\label{sec:experiment_results}

In this section, we present the experimental results and discuss their implications. We start with a brief introduction to our experimental setup in Section~\ref{subsec:experiment_results_setup_of_smac}. Then, we demonstrate that modeling a full distribution is beneficial to the performance of independent learners in Section~\ref{subsec:experiment_results_independent_learners}. Finally, we compare the performances of the CTDE baseline methods and their DFAC variants in Section~\ref{subsec:experiment_results_super_hard}.

\begin{figure*}[t]
\begin{tabular}{cc}
\begin{minipage}{0.7\textwidth} 
\includegraphics[width=\linewidth]{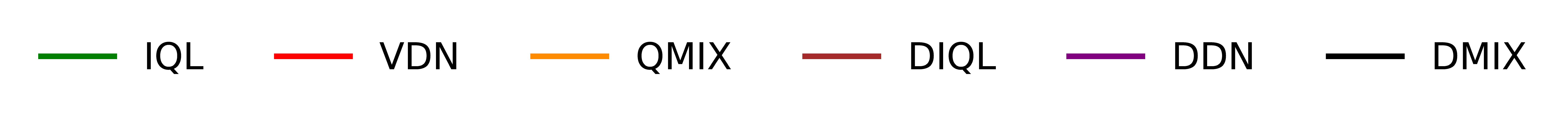}
\vspace{-1.5em}
\label{fig:win_rate_legend}
\end{minipage} \\
\begin{minipage}{0.19\textwidth} 
\includegraphics[width=\linewidth]{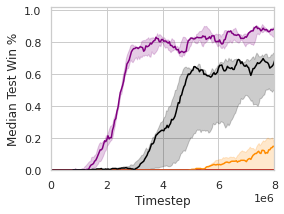}
\label{fig:win_rate_6h_vs_8z}
\vspace{-1.5em}
\subcaption{\texttt{6h\_vs\_8z}}
\end{minipage}
\begin{minipage}{0.19\textwidth} 
\includegraphics[width=\linewidth]{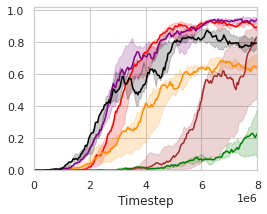}
\label{fig:win_rate_3s5z_vs_3s6z}
\vspace{-1.5em}
\subcaption{\texttt{3s5z\_vs\_3s6z}}
\end{minipage}
\begin{minipage}{0.19\textwidth} 
\includegraphics[width=\linewidth]{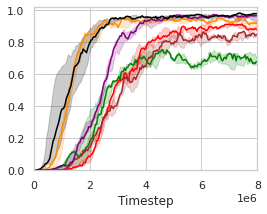}
\label{fig:win_rate_MMM2}
\vspace{-1.5em}
\subcaption{\texttt{MMM2}}
\end{minipage}
\begin{minipage}{0.19\textwidth} 
\includegraphics[width=\linewidth]{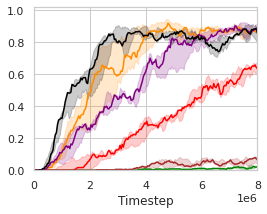}
\label{fig:win_rate_27m_vs_30m}
\vspace{-1.5em}
\subcaption{\texttt{27m\_vs\_30m}}
\end{minipage}
\begin{minipage}{0.19\textwidth} 
\includegraphics[width=\linewidth]{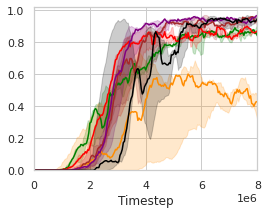}
\label{fig:win_rate_corridor}
\vspace{-1.5em}
\subcaption{\texttt{corridor}}
\end{minipage}
\end{tabular}
\caption{The win rate curves evaluated on the five \superhard{} maps in SMAC for different CTDE methods.}
\label{fig:smac_results_win_rate}
\end{figure*}

\begin{table*}[t]
\begin{minipage}{0.49\textwidth}
\footnotesize
\centering
\caption{The median win rate $\%$ of five independent test runs.}
\begin{tabular}{l|rrr|rrr}
\toprule
Map          & IQL   & VDN   & QMIX  & DIQL  & DDN            & DMIX \\
\midrule
\texttt{(a)} &  0.00 &  0.00 & 12.78 &  0.00 & \textbf{83.92} & 49.43 \\
\texttt{(b)} & 29.83 & 89.20 & 67.22 & 62.22 & \textbf{94.03} & 91.08 \\
\texttt{(c)} & 68.92 & 89.20 & 92.44 & 85.23 & \textbf{97.22} & 95.11 \\
\texttt{(d)} &  2.27 & 63.12 & 84.77 &  6.02 & \textbf{91.48} & 85.45 \\
\texttt{(e)} & 84.87 & 85.34 & 37.61 & 91.62 & \textbf{95.40} & 90.45 \\
\bottomrule
\end{tabular}
\label{table:smac_results_win_rate}
\small\\
\raggedright\ \ \ \ \textasteriskcentered\ Maps (a)-(e) correspond to the maps in Fig.~\ref{fig:smac_results_win_rate}.
\end{minipage}
~
\begin{minipage}{0.49\textwidth}
\small
\centering
\caption{The averaged scores of five independent test runs.}
\begin{tabular}{l|rrr|rrr}
\toprule
Map          & IQL   & VDN   & QMIX  & DIQL  & DDN            & DMIX \\
\midrule
\texttt{(a)} & 13.78 & 15.41 & 14.37 & 14.94 & \textbf{19.40} & 17.14 \\
\texttt{(b)} & 16.54 & 19.75 & 20.16 & 17.52 & \textbf{20.94} & 19.70 \\
\texttt{(c)} & 17.50 & 19.36 & 19.42 & 19.21 & \textbf{20.90} & 19.87 \\
\texttt{(d)} & 14.01 & 18.45 & 19.41 & 14.45 & \textbf{19.71} & 19.43 \\
\texttt{(e)} & 19.42 & 19.47 & 15.07 & 19.68 & \textbf{20.00} & 19.66 \\
\bottomrule
\end{tabular}
\label{table:smac_results_score}
\small\\
\raggedright\ \ \ \ \textasteriskcentered\ Maps (a)-(e) correspond to the maps in Fig.~\ref{fig:smac_results_win_rate}.
\end{minipage}
\end{table*}


\subsection{Experimental Setup}
\label{subsec:experiment_results_setup_of_smac}

\textbf{Environment.}
We verify the DFAC framework in the SMAC benchmark environments~\cite{Samvelyan2019SMAC} built on the popular real-time strategy game StarCraft II. Instead of playing the full game, SMAC is developed for evaluating the effectiveness of MARL micro-management algorithms. Each environment in SMAC contains two teams. One team is controlled by a decentralized MARL algorithm, with the policies of the agents conditioned on their local observation histories. The other team consists of enemy units controlled by the built-in game artificial intelligence based on carefully handcrafted heuristics, which is set to its highest difficulty equal to seven. The overall objective is to maximize the win rate for each battle scenario, where the rewards employed in our experiments follow the default settings of SMAC. The default settings use \textit{shaped rewards} based on the damage dealt, enemy units killed, and whether the RL agents win the battle. If there is no healing unit in the enemy team, the maximum return of an episode (i.e., the score) is $20$; otherwise, it may exceed $20$, since enemies may receive more damages after healing or being healed.

The environments in SMAC are categorized into three different levels of difficulties: \textit{Easy}, \textit{Hard}, and \textit{Super Hard} scenarios~\cite{Samvelyan2019SMAC}. In this paper, we focus on all \textit{Super Hard} scenarios including (a) \texttt{6h\_vs\_8z},  (b) \texttt{3s5z\_vs\_3s6z}, (c) \texttt{MMM2}, (d) \texttt{27m\_vs\_30m}, and (e) \texttt{corridor}, since these scenarios have not been properly addressed in the previous literature without the use of additional assumptions such as intrinsic reward signals~\cite{Du2019LIIR}, explicit communication channels~\cite{Zhang2019VBN,Wang2019NDQ}, common knowledge shared among the agents~\cite{De2019MACKRL,Wang2020ROMA}, and so on. Three of these scenarios have their maximum scores higher than $20$. In \texttt{3s5z\_vs\_3s6z}, the enemy \textit{Stalkers} have the ability to regenerate shields; in \texttt{MMM2}, the enemy \textit{Medivacs} can heal other units; in \texttt{corridor}, the enemy \textit{Zerglings} slowly regenerate their own health.

\textbf{Hyperparameters.}
For all of our experimental results, the training length is set to 8M timesteps, where the agents are evaluated every 40k timesteps with 32 independent runs. The curves presented in this section are generated based on five different random seeds. The solid lines represent the median win rate, while the shaded areas correspond to the $25^{\text{th}}$ to $75^{\text{th}}$ percentiles. For a better visualization, the presented curves are smoothed by a moving average filter with its window size set to 11. The detailed hyperparameter setups are provided in the supplementary material.

\textbf{Baselines.}
We select IQL, VDN, and QMIX as our baseline methods, and compare them with their distributional variants in our experiments. The configurations are optimized so as to provide the best performance for each of the methods considered.
Since we tuned the hyperparameters of the baselines, their performances are better than those reported in~\cite{Samvelyan2019SMAC}. The hyperparameter searching process is detailed in the supplementary material.


\subsection{Independent Learners}
\label{subsec:experiment_results_independent_learners}

In order to validate our assumption that distributional RL is beneficial to the MARL domain, we first employ the simplest training algorithm, IQL, and extend it to its distributional variant, called \diql{}. \diql{} is simply a modified IQL that uses IQN as its underlying RL algorithm without any additional modification or enhancements~\cite{Matignon2007Hysteretic,Lyu2020LikelihoodQuantile}.

From Figs.~\ref{fig:smac_results_win_rate}(a)-\ref{fig:smac_results_win_rate}(e) and Tables~\ref{table:smac_results_win_rate}-\ref{table:smac_results_score}, it is observed that \diql{} is superior to IQL even without utilizing any value function factorization methods. This validates that distributional RL has beneficial influences on MARL, when it is compared to RL approaches based only on expected values.


\subsection{Value Function Factorization Methods}
\label{subsec:experiment_results_super_hard}

In order to inspect the effectiveness and impacts of DFAC on learning curves, win rates, and scores, we next summarize the results of the baselines as well as their DFAC variants on the \textit{Super Hard} scenarios in Fig.~\ref{fig:smac_results_win_rate}(a)-(e) and Table~\ref{table:smac_results_win_rate}-\ref{table:smac_results_score}.

Fig.~\ref{fig:smac_results_win_rate}(a)-(e) plot the learning curves of the baselines and their DFAC variants, with the final win rates presented in Table~\ref{table:smac_results_win_rate}, and their final scores reported in Table~\ref{table:smac_results_score}. The win rates indicate how often do the player's team wins, while the scores represent how well do the player's team performs. Despite the fact that SMAC's objective is to maximize the win rate, the true optimization goal of MARL algorithms is the averaged score. In fact, these two metrics are not always positively correlated (e.g., VDN and QMIX in \texttt{6h\_vs\_8z} and \texttt{3s5z\_vs\_3s6z}, and QMIX and \dmix{} in \texttt{3s5z\_vs\_3s6z}).

It can be observed that the learning curves of \ddn{} and \dmix{} grow faster and achieve higher final win rates than their corresponding baselines. In the most difficult map: \texttt{6h\_vs\_8z}, most of the methods fail to learn an effective policy except for \ddn{} and \dmix{}. The evaluation results also show that \ddn{} and \dmix{} are capable of performing consistently well across all \superhard{} maps with high win rates. In addition to the win rates, Table~\ref{table:smac_results_score} further presents the final averaged scores achieved by each method, and provides deeper insights into the advantages of the DFAC framework by quantifying the performances of the learned policies of different methods.

The improvements in win rates and scores are due to the benefits offered by distributional RL~\cite{Lyle2019Comparative}, which enables the distributional variants to work more effectively in MARL environments. Moreover, the evaluation results reveal that \ddn{} performs especially well in most environments despite its simplicity. Further validations of \ddn{} and \dmix{} on our self-designed \ultrahard{} scenarios that are more difficult than \superhard{} scenarios can be found in our GitHub repository (\href{https://github.com/j3soon/dfac}{https://github.com/j3soon/dfac}), along with the gameplay recording videos.
\section{Conclusion}
\label{sec:conclusion}

In this paper, we provided a distributional perspective on value function factorization methods, and introduced a framework, called DFAC, for integrating distributional RL with MARL domains. We first proposed  DFAC based on a mean-shape decomposition procedure to ensure the Distributional IGM condition holds for all factorizable tasks. Then, we proposed the use of quantile mixture to implement the mean-shape decomposition in a computationally friendly manner. DFAC's ability to factorize the joint return distribution into individual utility distributions was demonstrated by a toy example. In order to validate the effectiveness of DFAC, we presented experimental results performed on all \superhard{} scenarios in SMAC for a number of MARL baseline methods as well as their DFAC variants. The results show that \ddn{} and \dmix{} outperform VDN and QMIX. DFAC can be extended to more value function factorization methods and offers an interesting research direction for future endeavors.

\section{Acknowledgements}
\label{sec:acknowledgements}

The authors acknowledge the support from NVIDIA Corporation and NVIDIA AI Technology Center (NVAITC). The authors thank Kuan-Yu Chang for his helpful critiques of this research work. The last author would like to thank the funding support from Ministry of Science and Technology (MOST) in Taiwan under grant nos. MOST 110-2636-E-007-010 and MOST 110-2634-F-007-019.


\bibliography{references}
\bibliographystyle{icml2021_style/icml2021}



\section*{Supplementary material}

# Supplementary Materials

Wei-Fang Sun[1,2]        Cheng-Kuang Lee[2]        Chun-Yi Lee[1]

[1]Department of Computer Science, National Tsing Hua University, Taiwan

[2]NVIDIA AI Technology Center, NVIDIA Corporation

## S1   Theorems and Proofs

In this section, we elaborate on the definitions, and provide the proofs of the theorems discussed in the main manuscript.

**Proposition 1.** *Monotonicity for utility distributions:*

$$Z_{jt}(\mathbf{h}, \mathbf{u}) = \Psi(Z_1(h_1, u_1), ..., Z_K(h_K, u_K)|s)$$
$$= M(Z_1(h_1, u_1), ..., Z_K(h_K, u_K)|s),$$

*where $M$ is a monotonic transformation that satisfies $\frac{\partial M}{\partial Q_k} \geq 0, \forall k \in \mathbb{K}$, is not a sufficient condition for DIGM, although the equality may hold for special cases of $M$ and $[Z_k(h_k, u_k)]_{k \in \mathbb{K}}$.*

*Proof.* We consider a degenerated case and prove the theorem by contradiction. Consider a case where there is only a single agent ($K = 1$), with a single fully observable state and an exponential transformation $M(Z_1(h_1, u_1)|s) = \exp(Z_1(h_1, u_1))$. The (joint) action space of this case consists of two (joint) actions: $\mathbb{U}_{jt} = \mathbb{U}_1 = \{u_1^*, u_1'\}$, where $u_1^*$ is the optimal action (with expected return 2) and $u_1'$ is the suboptimal action (with expected return 1.5). We define the probability mass function (PMF) of $Z_1(h_1, u_1^*)$ to be:

$$p(z) = \begin{cases} 1 & \text{if } z = 2 \\ 0 & \text{otherwise,} \end{cases}$$

and the PMF of $Z_1(h_1, u_1')$ to be:

$$p(z) = \begin{cases} 0.5 & \text{if } z = 0 \\ 0.5 & \text{if } z = 3 \\ 0 & \text{otherwise.} \end{cases}$$

By the definition above, we can calculate the followings:

$$\begin{cases} \mathbb{E}[Z_1(h_1, u_1^*)] = 1 \cdot 2 = 2 \\ \mathbb{E}[Z_1(h_1, u_1')] = 0.5 \cdot 0 + 0.5 \cdot 3 = 1.5 \\ \arg\max_{u_1} \mathbb{E}[Z_1(h_1, u_1)] = u_1^* \end{cases}$$

$$\begin{cases} \mathbb{E}[M(Z_1(h_1, u_1^*)|s)] = \mathbb{E}[\exp(Z_1(h_1, u_1^*))] = e^2 \approx 7.39 \\ \mathbb{E}[M(Z_1(h_1, u_1')|s)] = \mathbb{E}[\exp(Z_1(h_1, u_1'))] = 0.5 \cdot e^0 + 0.5 \cdot e^3 \approx 10.54 \\ \arg\max_{u_1} \mathbb{E}[M(Z_1(h_1, u_1)|s)] = u_1' \end{cases}$$

Assume, to the contrary, that *Monotonicity* for utility distributions is a sufficient condition for DIGM. By the definition of DIGM:

$$\arg\max_{\mathbf{u}} \mathbb{E}[Z_{\text{jt}}(\mathbf{h}, \mathbf{u})] = \left(\arg\max_{u_1} \mathbb{E}[Z_1(h_1, u_1)]\right)$$

$$\Rightarrow \arg\max_{u_1} \mathbb{E}[M(Z_1(h_1, u_1)|s)] = \arg\max_{u_1} \mathbb{E}[Z_1(h_1, u_1)]$$

$$\Rightarrow u_1' = u_1^* \ (\Rightarrow\Leftarrow \text{ contradiction}).$$

A contradiction occurs since $u_1' \neq u_1^*$, showing that *Monotonicity* is not a sufficient condition for DIGM. Since there exist a case where DIGM does not hold for $K = 1$, it certainly does not hold for all $K \in \mathbb{Z}$.

$\square$

**Theorem 1.** *Given a deterministic joint action-value function $Q_{\text{jt}}$, a stochastic joint action-value function $Z_{\text{jt}}$, and a factorization function $\Psi$ for deterministic utilities:*

$$Q_{\text{jt}}(\mathbf{h}, \mathbf{u}) = \Psi(Q_1(h_1, u_1), ..., Q_{\text{K}}(h_{\text{K}}, u_{\text{K}})|s),$$

*such that $[Q_k]_{k \in \mathbb{K}}$ satisfy IGM for $Q_{\text{jt}}$ under $\mathbf{h}$, the following distributional factorization:*

$$Z_{\text{jt}}(\mathbf{h}, \mathbf{u}) = \Psi(Z_1(h_1, u_1), ..., Z_{\text{K}}(h_{\text{K}}, u_{\text{K}})|s).$$

*is insufficient to guarantee that $[Z_k]_{k \in \mathbb{K}}$ satisfy DIGM for $Z_{\text{jt}}$ under $\mathbf{h}$.*

*Proof.* A contradiction is provided by Proposition 1.

$\square$

**Theorem 2** (DFAC Theorem)**.** *Given a deterministic joint action-value function $Q_{\text{jt}}$, a stochastic joint action-value function $Z_{\text{jt}}$, and a factorization function $\Psi$ for deterministic utilities:*

$$Q_{\text{jt}}(\mathbf{h}, \mathbf{u}) = \Psi(Q_1(h_1, u_1), ..., Q_{\text{K}}(h_{\text{K}}, u_{\text{K}})|s),$$

*such that $[Q_k]_{k \in \mathbb{K}}$ satisfy IGM for $Q_{\text{jt}}$ under $\mathbf{h}$, by Mean-Shape Decomposition, the following distributional factorization:*

$$
\begin{aligned}
Z_{\text{jt}}(\mathbf{h}, \mathbf{u}) &= \mathbb{E}[Z_{\text{jt}}(\mathbf{h}, \mathbf{u})] + (Z_{\text{jt}}(\mathbf{h}, \mathbf{u}) - \mathbb{E}[Z_{\text{jt}}(\mathbf{h}, \mathbf{u})]) \\
&= Z_{\text{mean}}(\mathbf{h}, \mathbf{u}) + Z_{\text{shape}}(\mathbf{h}, \mathbf{u}) \\
&= \Psi(Q_1(h_1, u_1), ..., Q_{\text{K}}(h_{\text{K}}, u_{\text{K}})|s) \\
&\quad + \Phi(Z_1(h_1, u_1), ..., Z_{\text{K}}(h_{\text{K}}, u_{\text{K}})|s).
\end{aligned}
$$

*is sufficient to guarantee that $[Z_k]_{k \in \mathbb{K}}$ satisfy DIGM for $Z_{\text{jt}}$ under $\mathbf{h}$, where $\text{Var}(\Psi) = 0$ and $\mathbb{E}[\Phi] = 0$.*

*Proof.* By mean-shape decomposition:

$$\arg\max_{\mathbf{u}} \{\mathbb{E}[Z_{\mathrm{jt}}(\mathbf{h}, \mathbf{u})]\}$$

$$= \arg\max_{\mathbf{u}} \{\mathbb{E}[Z_{\mathrm{mean}}(\mathbf{h}, \mathbf{u}) + Z_{\mathrm{shape}}(\mathbf{h}, \mathbf{u})]\}$$

$$= \arg\max_{\mathbf{u}} \{\mathbb{E}[Z_{\mathrm{mean}}(\mathbf{h}, \mathbf{u})] + \mathbb{E}[Z_{\mathrm{shape}}(\mathbf{h}, \mathbf{u})]\}$$

$$= \arg\max_{\mathbf{u}} \{\mathbb{E}[\Psi(Q_1(h_1, u_1), ..., Q_{\mathrm{K}}(h_{\mathrm{K}}, u_{\mathrm{K}})|s)]$$
$$+ \mathbb{E}[\Phi(Z_1(h_1, u_1), ..., Z_{\mathrm{K}}(h_{\mathrm{K}}, u_{\mathrm{K}})|s)]\}$$

$$= \arg\max_{\mathbf{u}} \{\Psi(Q_1(h_1, u_1), ..., Q_{\mathrm{K}}(h_{\mathrm{K}}, u_{\mathrm{K}})|s) + 0\}$$

$$= \arg\max_{\mathbf{u}} \{\Psi(Q_1(h_1, u_1), ..., Q_{\mathrm{K}}(h_{\mathrm{K}}, u_{\mathrm{K}})|s)\}$$

$$= \begin{pmatrix} \arg\max_{u_1} Q_1(h_1, u_1) \\ \vdots \\ \arg\max_{u_{\mathrm{K}}} Q_{\mathrm{K}}(h_{\mathrm{K}}, u_{\mathrm{K}}) \end{pmatrix}$$

$$\Rightarrow \arg\max_{\mathbf{u}} \mathbb{E}[Z_{\mathrm{jt}}(\mathbf{h}, \mathbf{u})] = \begin{pmatrix} \arg\max_{u_1} \mathbb{E}[Z_1(h_1, u_1)] \\ \vdots \\ \arg\max_{u_{\mathrm{K}}} \mathbb{E}[Z_{\mathrm{K}}(h_{\mathrm{K}}, u_{\mathrm{K}})] \end{pmatrix}.$$

The equations above show that $[Z_k]_{k \in \mathbb{K}}$ satisfy *DIGM* for $Z_{\mathrm{jt}}$ under $\mathbf{h}$. $\qquad\square$

**Theorem 3.** *Given a quantile mixture:*

$$F^{-1}(\omega) = \sum_{k=1}^{\mathrm{K}} \beta_k F_k^{-1}(\omega)$$

*with* K *components* $[F_k^{-1}]_{k \in \mathbb{K}}$ *and non-negative model parameters* $[\beta_k]_{k \in \mathbb{K}}$. *There exist a set of random variables* $Z = F^{-1}(\tau)$ *and* $[Z_k = F_k^{-1}(\tau)]_{k \in \mathbb{K}}$ *corresponding to the quantile functions* $F^{-1}$ *and* $[F_k^{-1}]_{k \in \mathbb{K}}$, *respectively, where* $\tau$ *is a random variable uniformly distributed on* $[0, 1]$, *with the following relationship:*

$$Z \stackrel{D}{=} \sum_{k \in \mathbb{K}} \beta_k Z_k.$$

*Proof.* We first prove the case for a quantile mixture with K = 2 components, and then generalize it to all K $\in \mathbb{Z}$. For K = 2, the quantile mixture is simplified as follows:

$$F^{-1}(\tau) = \beta_1 F_1^{-1}(\tau) + \beta_2 F_2^{-1}(\tau)$$

For notational simplicity, let $X = \beta_1 Z_1$, $Y = \beta_2 Z_2$, and $\tau$ is a latent variable shared among the random variables $X$, $Y$, and $Z$. The corresponding CDFs of the random variables $X$, $Y$, and $Z$ are $F_X$, $F_Y$, and $F_Z$, respectively, with $X(\tau) = F_X^{-1}(\tau)$, $Y(\tau) = F_Y^{-1}(\tau)$, and $Z(\tau) = F_Z^{-1}(\tau)$. Under this notation, the above equation can be re-written as:

$$F_Z^{-1}(\tau) = F_X^{-1}(\tau) + F_Y^{-1}(\tau).$$

The goal is to prove that there exist random variables $(X, Y, Z)$ such that the following holds:

$$Z \stackrel{D}{=} X + Y$$

By the definition of the CDF of $X + Y$, the following holds:

$$
\begin{aligned}
& F_{X+Y}(z), \forall z \in \mathbb{R} \\
=\ & \Pr(X + Y \leq z), \forall z \in \mathbb{R} \\
=\ & \Pr(\{\tau \in [0,1] : X(\tau) + Y(\tau) \leq z\}), \forall z \in \mathbb{R} \\
=\ & \Pr(\{\tau \in [0,1] : F_X^{-1}(\tau) + F_Y^{-1}(\tau) \leq z\}), \forall z \in \mathbb{R} \\
=\ & \sup\{\tau \in [0,1] : F_X^{-1}(\tau) + F_Y^{-1}(\tau) \leq z\}, \forall z \in \mathbb{R} \\
=\ & \inf\{\tau \in [0,1] : z \leq F_X^{-1}(\tau) + F_Y^{-1}(\tau)\}, \forall z \in \mathbb{R} \\
=\ & \inf\{\tau \in [0,1] : z \leq F_Z^{-1}(\tau)\}, \forall z \in \mathbb{R} \\
=\ & F_Z(z), \forall z \in \mathbb{R}. \\
\Rightarrow\ & Z \stackrel{D}{=} X + Y.
\end{aligned}
$$

The proof for quantile mixtures with two components can be iteratively applied to quantile mixtures with K $\in \mathbb{Z}$ components.

$\square$

# S2  Hyperparameters and Settings

## S2.1  Stochastic Two Step Game

In the stochastic two step game described in Section 4, each agent is implemented as an IQN with two hidden layers comprised of $64$ units and $512$ units, respectively, with a ReLU nonlinearity at the end of each layer. We optimize the IQNs with $N = N' = 32$ quantile samples, where each of them is encoded into a $64$-dimensional intermediate embedding and projected to a $512$-dimensional quantile embedding by a single hidden layer. Each agent performs independent $\epsilon$-greedy action selection, with full exploration (i.e., $\epsilon = 1$). We set the discount factor $\gamma$ to $0.99$. The replay buffer contains the state-action pairs of the latest $2k$ episodes, from which we uniformly sample a batch of size $512$ for training. The target network is updated every 100 episodes. The optimizer is set to Adam, in which its learning rate is set to $1 \times 10^{-4}$. We train for $20k$ timesteps ($10k$ episodes). All agent networks share parameters, and the one-hot encoded agent id ($[1\ 0]^T$ for agent 1 and $[0\ 1]^T$ for agent 2) is concatenated to each agent's observation. We do not pass the previous actions taken by the agents as their inputs. Each agent receives the full state as its input. For *DMIX*, we use a mixing network with $8$ units.

Each state is one-hot encoded. The starting state for the first timestep is State 1 (one-hot: $[1\ 0\ 0]^T$). At State 1, if Agent 1 selects Action A, the agents transit to State 2A (one-hot: $[0\ 1\ 0]^T$). On the other hand, if agent 1 selects Action B, the agents transit to State 2B (one-hot: $[0\ 0\ 1]^T$).

## S2.2  SMAC

We tuned the hyperparameters of both the baselines and their distributional variants by selecting their hidden layer sizes from $\{32, 64, 128, 256, 512\}$ and choose the best ones. The quantile samples of *DIQL* and *DDN*

Table S1: A summary of the optimal hidden state sizes of the baseline methods and their distributional variants.

| Maps | IQL | VDN | QMIX | QR-MIX | DIQL | DDN | DMIX |
|------|-----|-----|------|--------|------|-----|------|
| `3s5z_vs_3s6z` | 512 | 128 | 128 | 128 | 256 | 512 | 256 |
| `6h_vs_8z` | 128 | 128 | 256 | 256 | 512 | 512 | 256 |
| MMM2 | 256 | 64 | 64 | 128 | 512 | 512 | 256 |
| `27m_vs_30m` | 256 | 64 | 64 | 64 | 512 | 128 | 128 |
| `corridor` | 256 | 128 | 256 | 64 | 512 | 128 | 64 |

Table S2: The detailed settings of the *Super Hard* maps.

| Difficulty | Map | Player's Team | Enemy's Team |
|------------|-----|---------------|--------------|
| *Super Hard* | `6h_vs_8z` | 6 Hydralisks | 8 Zealots |
| | `3s5z_vs_3s6z` | 3 Stalkers & 5 Zealots | 3 Stalkers & 6 Zealots |
| | MMM2 | 7 Marines, 2 Marauders & 1 Medivac | 8 Marines, 3 Marauders & 1 Medivac |
| | `27m_vs_30m` | 27 Marines | 30 Marines |
| | `corridor` | 6 Zealots | 24 Zerglings |

are simply set to $N = N' = 1$, since they do not require the calculation of the expected value during the optimization process. As for *DMIX*, the numbers of quantile samples are set to $N = N' = 8$ as in [1]. The optimizers follow those used in DQN and IQN. All of the other hyperparameters follow those used in SMAC. Table S1 lists the hyperparameters adopted for the baselines and their distributional variants. The StarCraft version we used is 4.10.



\end{document}